# Experimental Verification of Illusion Optics with Optical Null Media

Mohammad Mehdi Sadeghi [1, *] and Hayrettin Odabasi[2]

[1] *Department of Physics, Jahrom University, Jahrom 74137-66171, Iran*

[2] *Electrical and Electronics Engineering, Eskisehir Osmangazi University, Eskisehir, Turkey*

*Corresponding Author: sadeghi@jahromu.ac.ir*

**Abstract**: We experimentally show the results of the illusion device designed by the transformation optics (TO) technique. The coordinate transformation maps a zero thickness in virtual space to a finite region in physical space, yielding extreme material properties called optical null media (ONM). As a result, an object placed inside the core media will look larger to an outside observer. ONM media is realized using metal-dielectric layered structures in cylindrical coordinates. Also closed-form solutions for the scattering field from a cylindrical object are given in detail to prove illusion effect analytically. ONM media is realized using air and iron sheets in radial structure using effective medium theory. We experimentally fabricate the ONM and show the scattering effect using the ONM. For achieving illusion validation, a metallic pipe has been inserted into the center of the core medium made of Plexiglas and surrounded by ONM.

## 1- Introduction

Relying on the form invariance of Maxwell's equations [1-3], transformation optics (TO) [4-7] provides a methodology to control and manipulate the electromagnetic waves in order to design novel devices such as invisibility cloaks [4,5,8-10], optical black holes [11-15], negative index lenses [6,7,16-18], concentrators [19,20], rotators [21-25], PEC reshaper [26,27], super scatterer [28-31], reflectionless waveguide bends [32,33], waveguide miniaturization [34,35] including many other devices. These devices, typically, are restricted by their anisotropic characteristics, narrow bandwidth, and high losses associated with resonant metamaterial elements. With the advance in metamaterial technology, however, some of these difficulties can be elevated, and these devices can be realized with increased capabilities [36-41].

One media that has recently attracted great attention is the optical null media (ONM) [42-45]. ONM media can be derived by transforming zero thickness into a finite region. As a result, the derived transformation media has extreme anisotropic material properties (i.e., $[\varepsilon], [\mu] = \{\infty, 0, 0\}$). ONM has two important characteristics. First, phase accumulation inside the transformation media is zero; thus, ONM nullifies the occupying space. Second, the wave propagation is limited to one particular direction only. These properties are particularly used to realize interesting surface transformation devices [46,47]. Many useful devices have been designed using ONM, including hyper-lens [48], magnetic hoses [49], and field concentrators [50]. Such anisotropic material properties can be realized using metal-dielectric layered structures [51-52].

Extreme material properties, for instance, ONM, can be obtained by appropriately arranging the ratio of metallic and dielectric layer thickness.

In this paper, we design an illusion optics device using the TO approach. In the coordinate transformation, zero thickness in the virtual space is mapped to a finite thickness in physical space, yielding extreme material properties. Any object placed inside the transformation device's core region will look larger from an outside observer. We provide analytical calculations for the scattered field and show the equivalence between the original and equivalent problems. ONM is realized using air and iron metal layers in cylindrical coordinates. We show that the proposed structure approximates the effective medium parameters closely. We have also experimentally fabricated the transformation media and showed the illusion effect. The proposed device is realized using isotropic air and iron to realize ONM and does not require resonant metamaterial structures.

## 2- Optical Null Media for Illusion Optics

Here, we first study the derivation of the ONM media. Fig. 1(a,b) shows the virtual and physical space, respectively, where almost zero thickness region ($\delta$) in virtual space is mapped to a finite thickness ($\Delta$) in physical space. The corresponding coordinate transformation from physical space to virtual space is given in Eq. (1).

$$r = \begin{cases} \frac{R_2 - \Delta}{R_2 - \delta} r' & r \leq R_1 \\ \frac{\Delta}{\delta} r' - \left(\frac{\Delta}{\delta} - 1\right) R_2 & R_1 \leq r \leq R_2 \\ r' & r \geq R_2 \end{cases} \quad (1)$$

where $r$ and $r'$ represent virtual and physical space, respectively. In this relation, $\delta$ denotes the region between $R_2$ and $(R_2 - \delta)$ is mapped to a finite region between $R_2$ and $R_1$ denoted with $\Delta$. Using the TO approach, the corresponding material properties can be derived as follows,

$$[\varepsilon] = [\mu] = \begin{cases} \text{diag}\left(1,1,\left(\frac{R_2-\delta}{R_2-\Delta}\right)^2\right) & r \leq R_1 \\ \text{diag}\left(P,\frac{1}{P},\left(\frac{\delta}{\Delta}\right)^2 P\right) & R_1 \leq r \leq R_2 \end{cases} \quad (2)$$

where $P = 1 + \left(\frac{\Delta}{\delta} + 1\right)\frac{R_2}{r}$. Note that the corresponding material parameters are extremely large in the radial direction and near zero for other directions. In extreme scenario where $\delta \to 0$, the above tensors become,

$$[\varepsilon] = [\mu] = \begin{cases} \text{diag}\left(1,1,\left(\frac{R_2}{R_1}\right)^2\right) & r \leq R_1 \\ \text{diag}(\infty, 0, 0) & R_1 \leq r \leq R_2 \end{cases} \quad (3)$$

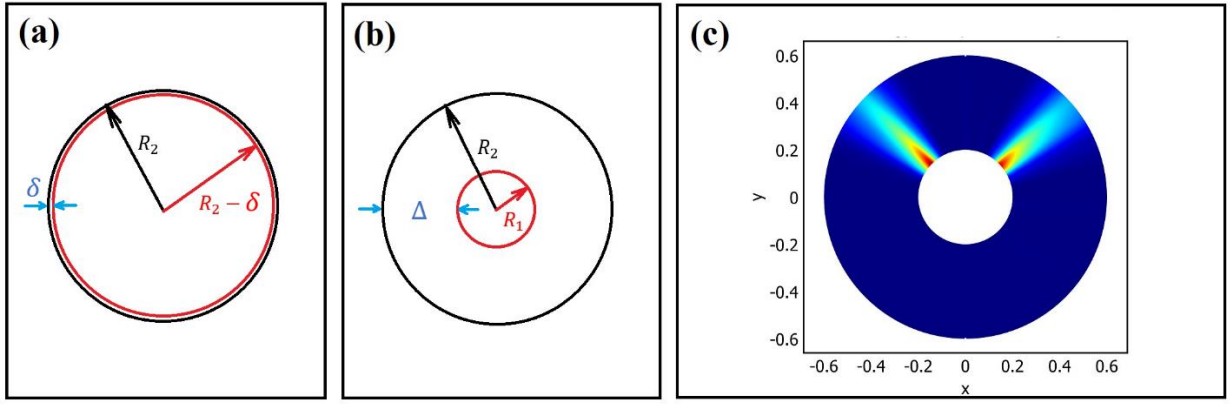

Figure 1. (a) and (b) show the transformation relation between virtual space and real space, respectively. (c) Numerical simulation of energy density of Gaussian beam in 2D ONM.

The transformed media between $R_2$ and $R_1$ is called optical null media (ONM). As mentioned before, one interesting property of ONM is that the wave propagation happens in one radial direction. This behavior can be seen from Fig.1(c), where a Gaussian beam is incident on a cylindrical ONM. This behavior can also be concluded from the dispersion relation, i.e., $\frac{k_r^2}{\varepsilon_\varphi} + \frac{k_\varphi^2}{\varepsilon_r} = \omega^2$. With extremely large values of $\varepsilon_r$, the wave propagation is confined in one direction. Such transformed media can be used for illusion purposes. For instance, any object placed in the core region ($r' \leq R_1$) will be seen as larger object by outside observer. The enlargement ratio depends on the values of $R_1$ and $R_2$. Note also that when the red line in Fig. 1(a) is outside of $R_2$ (i.e., $R_2 + \delta$), the transformation yields negative material properties. For instance, a superscatter device is designed by using such coordinate transformation and complementary media [29].

Assume that a cylindrical object with a radius $d$ is placed inside the core region (($r' \leq R_1$)). The scattering problem can be solved to find field solutions in four different regions. Considering a $TM_z$ plane wave is incident from outside with the wave vector $K_0$, the field solutions in all regions can be written as,

$$H_z = \begin{cases} \sum_n [i^{-n} J_n(K_0 r) + b_n H_n^{(1)}(K_0 r)] e^{in\theta} & r \geq R_2 \\ g J_0(K_0 r) + h Y_0(K_0 r) & R_1 \leq r \leq R_2 \\ \sum \left[ c_n J_n\left(\frac{R_2}{R_1} K_0 r\right) + d_n H_n^{(1)}\left(\frac{R_2}{R_1} K_0 r\right) \right] e^{in\theta} & d \leq r \leq R_1 \\ \sum_n f_n J_n(m \frac{R_2}{R_1} K_0 r) e^{in\theta} & r \leq d \end{cases} \quad (4)$$

where $J_n, H_n, Y_n$ are Bessel functions. $g, h, b_n, c_n, d_n, f_n$, are the scattering coefficients in each region. If the object in core region ($r \leq d$) was made of dielectric $f_n$ is unknown coefficient of wave and $m$ is refractive index of dielectric cylinder, whereas if the sample is a metallic cylinder then $f_n = 0$. Using the boundary condition for electric and magnetic fields at all interfaces, the scattering coefficient outside of the device can be calculated as,

$$b_n = i^{-n} \frac{[A_n J_0(K_0 R_2) - B_n Y_0(K_0 R_2)] J_n'(K_0 R_2) + [B_n Y_0'(K_0 R_2) - A_n J_0'(K_0 R_2)] J_n(K_0 R_2)}{[A_n J_0'(K_0 R_2) - B_n Y_0'(K_0 R_2)] H_n(K_0 R_2) + [B_n Y_0(K_0 R_2) - A_n J_0(K_0 R_2)] H_n'(K_0 R_2)} \quad (5)$$

where $A_n = M_n(K_0R_2)Y_0'(K_0R_1) - M_n'(K_0R_2)Y_0(K_0R_1)$, $B_n = M_n(K_0b)J_0'(K_0R_1) - M_n'(K_0R_2)J_0(K_0R_1)$ and $M_n(K_0R_2) = \beta_1 J_n(K_0R_2) + \beta_2 H_n(K_0R_2)$. With that the field distribution, outside of the ONM can be calculated. In order to compare the scattering results with a larger transformed bare cylindrical object, the scattering coefficients can be calculated similar to previous calculations. In this case, according to TO, the radius of the object becomes as $d' = \frac{R_2}{R_1}d$. In the equivalent problem, the field solutions can be expressed as,

$$H_z = \begin{cases} \sum \left[i^{-n}J_n(K_0r) + b_n H_n^{(1)}(K_0r)\right]e^{in\theta} & d' \leq r \\ \sum [f_n J_n(m'K_0r)] e^{in\theta} & r \leq d' \end{cases} \quad (6)$$

Using boundary condition for electric and magnetic fields at $r = d'$, we can calculate the scattering coefficient outside the metal cylinder, in region $r > d'$,

$$b_n = i^{-n} \frac{J_n'(k_0R_2)J_n(mk_0R_2) - J_n(k_0R_2)J_n'(mk_0R_2)}{H_n(k_0R_2)J_n'(mk_0R_2) - H_n'(k_0R_2)J_n(mk_0R_2)} \quad \ldots\ldots(7)$$

Fig.2 shows the calculated fields for the two cases. In first case, ONM media is placed between $R_1 = 0.1m$ and $R_2 = 0.3m$ the metallic cylinder with radius $d = 0.5m$ is inserted into the core region. In the second case, a metallic cylinder with radius $d' = 0.15$ is placed in free space. In these numerical calculations, the frequency is chosen as 1 GHz. As can be seen from both figures the scattered field outside of region $R_2$, outside of yellow circle, is same for both configurations.

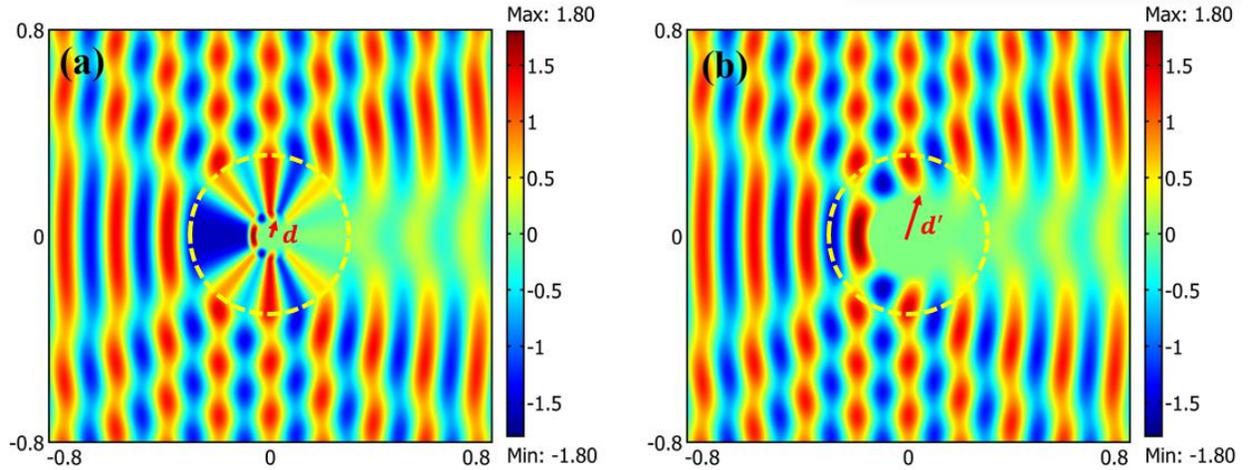

Figure 2 (a) Magnetic field pattern of ONM including a metallic pipe in core medium with radius of d (b) magnetic field pattern of bare transformed metallic cylinder with radius of $d'$.

## 3- Experimental Results

In this section, we propose a radially metal-dielectric structure for the realization of ONM, as illustrated in Fig. 3(a). The structure composed of slices made by air and iron as depicted in Fig. 3(a) labeled with letter A and B, respectively. The effective material properties for this structure can be obtained by effective medium theory (EFT) [51-53]. Assuming that the filling ratio of air and iron is given as $f_A = 0.903$ and $f_B = 0.097$, where the thickness of the iron is set as 0.4 mm. The dielectric constants are given as $\varepsilon_A = 1$ and $\varepsilon_A = 100000$. With that effective parameters can be written,

$$\varepsilon_r = f_A \varepsilon_A + f_B \varepsilon_B \tag{8}$$

$$\varepsilon_\theta^{-1} = f_A/\varepsilon_A + f_B/\varepsilon_B \tag{9}$$

Where for the above ratios and values we find $\varepsilon_r = 9644$ and $\varepsilon_\theta = 1.1$. In the simulations, we used 0.4 mm thickness for the iron sheet. Fig.3(b) show the Gaussian beam propagation for the effective medium parameters. Note that although the $\varepsilon_\theta$ is not zero with the proposed structure, extreme anisotropy in material properties mimics the ONM characteristics closely. The proposed structure can be used to realize ONM.

It is noteworthy that in this structure we use reduced parameters in core medium instead of perfect parameters of eq. (3). In fact, by keeping the refractive index constant $n_r = \varepsilon_\theta \mu_z$ and $n_\theta = \varepsilon_r \mu_z$, we can introduce reduced parameters instead of perfect one. For the core medium we have Plexiglas with isotropic $\varepsilon_P = 2.7$ and $\mu = 1$. In this case, although in the end the refractive index does not change, comparing to perfect device, but it causes impedance mismatch in boundary of core medium and ultimately creates a considerable scattering wave that affects the illusion. To reduce the effects of this problem, we have to use fabry-perot Fabric frequencies to minimize this scattering as much as possible [45].

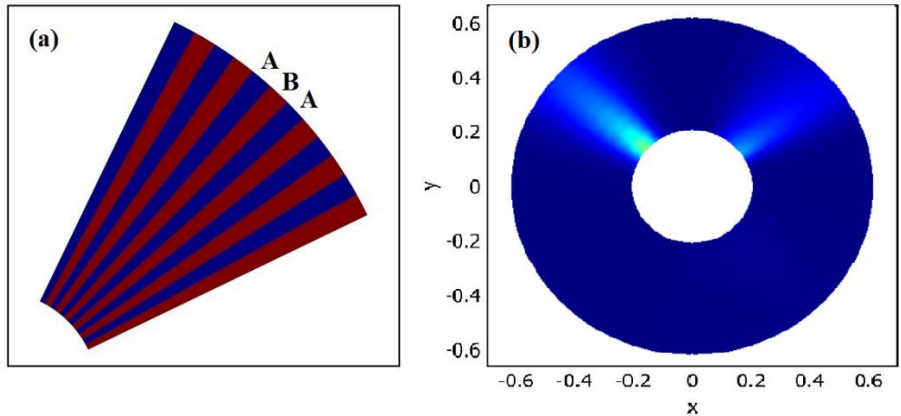

Figure 3. (a)periodic structure in radial alternative form comprised of air and metallic slits. (b) electric energy density in medium with have $\varepsilon_r = 9644$ and $\varepsilon_\varphi = 1.1$.

Experimental configurations are shown in Fig. 4. Fig 4(a) shows the Plexiglas cylindrical core medium with dielectric constant of $\varepsilon_P = 2.7$ and radius of 50mm with hole in center of the core medium for inserting the metallic pipe, i.e., $R_1$=50mm and $d = 12.5$mm. Figure 4(b) shows the metallic cylinder with radius of $d = 12.5$ mm. is placed inside the Plexiglas core. Finally, Fig. 4(c) shows the ONM outside of the Plexiglas core is realized with air/iron layered structure where $32 \times 500$ mm iron sheets with 0.4 mm thickness are used to realize ONM. A total of 100 iron sheets is used to realize ONM.

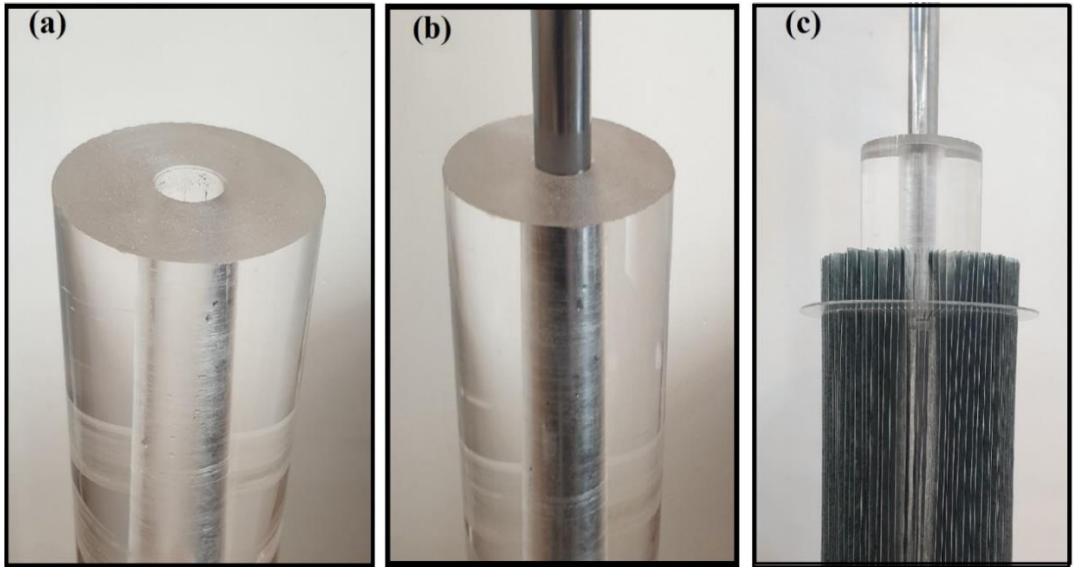

Figure 4 (a) core region of device made by plexiglass with permittivity $\varepsilon = 2.7$ and a central hole with radius d=12.5mm (b) core region of device including a metallic pipe in the central hole(c) fabricated device with annual slits media

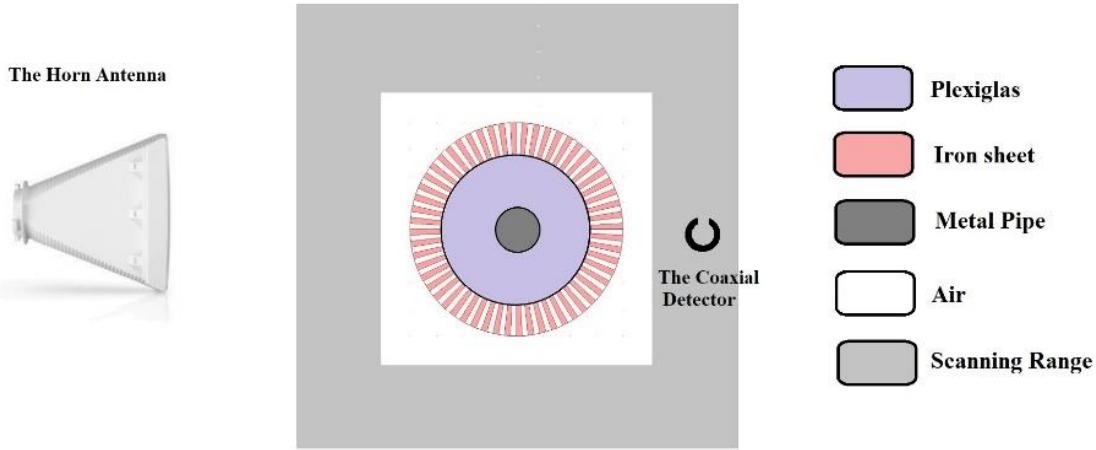

Figure 5. Schematic experimental set up.

The experiment set up is depicted in Fig. 5 where the sample is placed vertically in front horn-antenna that is located 100 cm away from sample. The horn antennas is arranged so that the $TM_z$ fields are incident on the sample. Then the scattered field is detected using a split ring detecting antenna with circular shape with radius of 4mm and a split of 1mm. The scanning area is chosen between the 300x300 mm and 200x200 mm square region as shown with gray color in Fig. 5. The spatial resolution is chosen as 2mm in both directions. Two experiment is performed in order to show the illusion effect. In the first experiment, a metallic cylinder with radius $d = 12.5$mm is placed inside a core medium covered with ONM. In the second experiment we measured the scattering fields from a metallic cylinder with a radius of $d' = 20.5$ is placed in free space. Our goal is to show that ONM achieves an optical illusion of size increase of an object placed inside the transformation media.

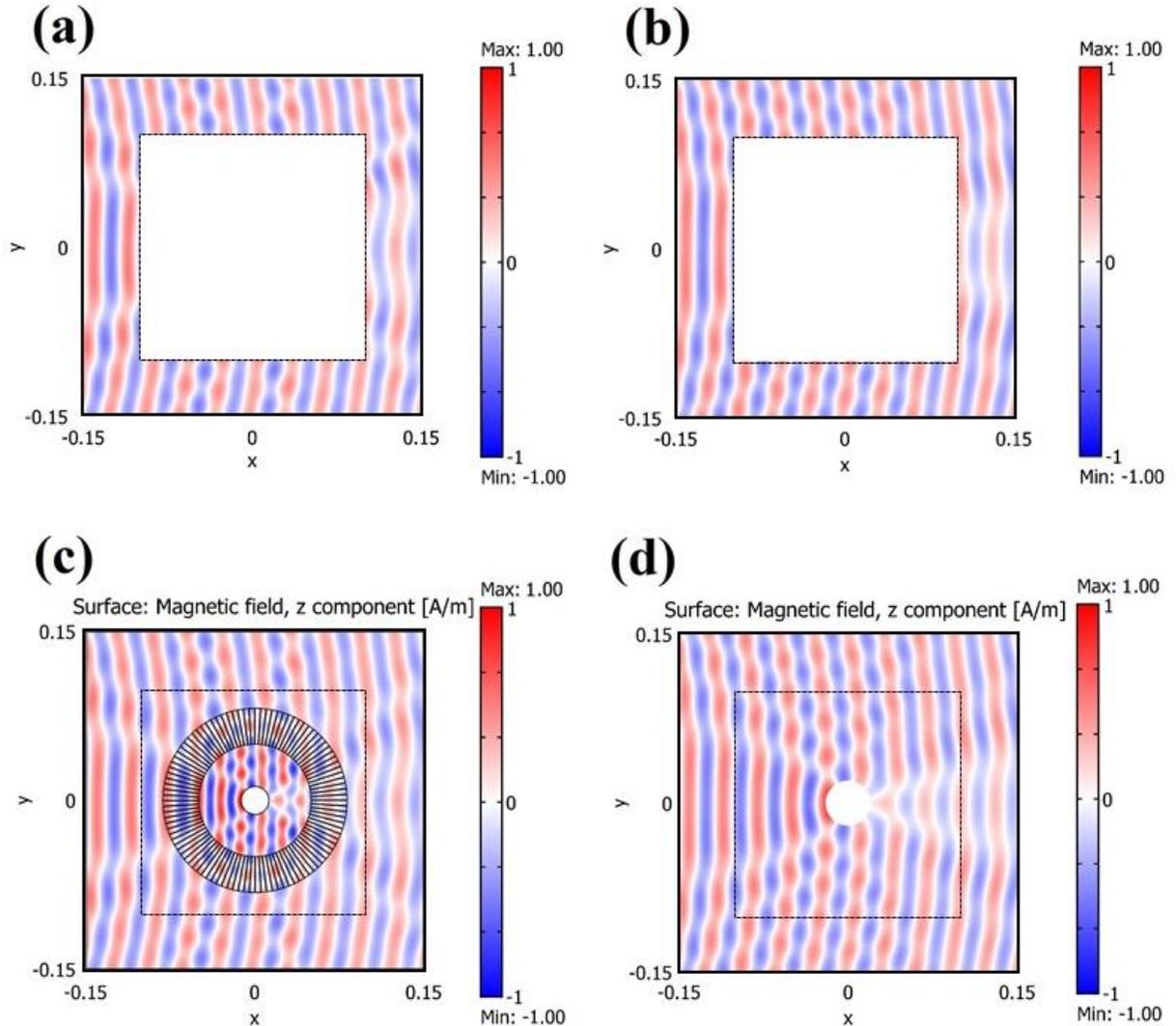

Figure 6 (a) Experimental measuring of Hz-field of the ONM media including metallic cylinder with radius $d = 12.5mm$ and working frequency 9.86 GHz (b) Experimental measuring of Hz-field of the bare metallic cylinder with radius $d' = 20.5mm$ and working frequency 9.86 GHz. (c) the Comsol simulation results of ONM case using layered iron/air structure. (d) ) the Comsol simulation results of the scattering results for transformed bare metallic cylinder with radius $d' = 20.5mm$.

Fig 6 shows the experimental scattering results (real part of $H_z$). Fig. 6(a) shows the experiment results of ONM case using layered iron/air structure. Fig. 6(b) shows the scattering results for larger transformed bare metallic cylinder. The experiment is performed at 9.86 GHz. Also Fig 6c shows the Comsol simulation results of ONM case using layered iron/air structure. fig 6(d) shows the comsol simulation of the scattering results for larger bare metallic cylinder with radius $d' = 20.5mm$.

## 4- Conclusion

This paper presents an experimental realization of an illusion optics device where a metallic object placed inside the core transformation media is enlarged via ONM. The scattering coefficients for the ONM with an object inserted inside the core medium are calculated theoretically and compared with a transformed bare object result. ONM is realized with an iron/air radial structure. The proposed structure is fabricated and experimentally verified the illusion effect. Both analytical and experimental results agree well.

## 5- References